\documentclass{aastex63}
\usepackage{amsmath, amsthm}
\usepackage{float}
\usepackage[caption=false]{subfig}

\newcommand\symphony{{\tt symphony}}
\newcommand\harmony{{\tt harmony}}
\newcommand\kdf{{$\kappa$}}
\newcommand{\sign}{\text{sgn}}
\usepackage[displaymath, mathlines]{lineno}
\defcitealias{PaperVII}{EHTC~VII}
\defcitealias{PaperVIII}{EHTC~VIII}

\submitjournal{ApJ}

\shorttitle{Updated Transfer Coefficients}
\shortauthors{Marszewski et al.}

\graphicspath{{./}{figures/}}

\usepackage{natbib}
\begin{document}

\title{Updated Transfer Coefficients for Magnetized Plasmas}

\author{Andrew Marszewski}
\affiliation{Department of Physics, University of Illinois, 1110 West Green Street, Urbana, IL 61801, USA}
\email{agm5@illinois.edu}

\author{Ben S. Prather}
\affiliation{Department of Physics, University of Illinois, 1110 West Green Street, Urbana, IL 61801, USA}

\author{Abhishek V. Joshi}
\affiliation{Department of Physics, University of Illinois, 1110 West Green Street, Urbana, IL 61801, USA}

\author{Alex Pandya}
\affiliation{Department of Physics, Princeton University, Princeton, New Jersey 08544, USA}
\email{apandya@princeton.edu}

\author{Charles F. Gammie}
\affiliation{Department of Physics, University of Illinois, 1110 West Green Street, Urbana, IL 61801, USA}
\affiliation{Department of Astronomy, University of Illinois, 1002 West Green Street, Urbana, IL 61801, USA}
\email{gammie@illinois.edu}

\begin{abstract}

Accurate radiative transfer coefficients (emissivities, absorptivities, and rotativities) are needed for modeling radiation from relativistically hot, magnetized plasmas such as those found in Event Horizon Telescope sources.  Here we review, update, and correct earlier work on radiative transfer coefficients.  We also describe an improved method for numerically evaluating rotativities and provide convenient fitting formulae for the relativistic \kdf{} distribution of electron energies.   

\end{abstract}

\section{Introduction} \label{sec:intro}
We are interested in modelling polarized radiative transfer within the magnetized plasma around black holes.  Evaluations of transfer coefficients date back as far as \citet{westfold_polarization_1959}.  A summary of early work on emission and absorption coefficients is given by \citet{ginzburg_cosmic_1965} \textcolor{black}{(see also \citet{Fleishman_2010})}.  Recently, \citet{dexter_public_2016} and \citet{moscibrodzka_ipole_2018} have used polarized transfer coefficients to model emission from the accretion flows of black holes in their respective ray-tracing codes, \texttt{GRTRANS} and \texttt{ipole}, and these have been used in modeling Event Horizon Telescope observations of M87 (\citetalias{PaperVII}, \citetalias{PaperVIII}).

Polarized intensity is described by the Stokes vector $I_S = \{I, Q, U,V\}$, where Stokes I is the total intensity, Q and U describe linear polarization, and V describes circular polarization.  Polarized radiative transfer can be described by a set of emission, absorption, and Faraday mixing coefficients in the Stokes basis.  The radiative transfer equation in the Stokes basis is
\begin{linenomath}
\begin{equation}
    \frac{d}{ds}I_S = j_S-M_{ST}I_T,
\end{equation}
\end{linenomath}
where $j_S$ is a vector containing the emission coefficients in the Stokes basis, \textcolor{black}{$ds$ is the path length, $S$ and $T$ are indices of the Stokes vector,} and $M_{ST}$ is the Mueller matrix:
\begin{linenomath}
\begin{equation}
    M_{ST} \equiv \begin{pmatrix} \alpha_I & \alpha_Q & \alpha_U & \alpha_V\\ \alpha_Q & \alpha_I & \rho_V & -\rho_U \\ 
    \alpha_U & -\rho_V & \alpha_I & \rho_Q\\
    \alpha_V & \rho_U & -\rho_Q & \alpha_I  \end{pmatrix}.
\end{equation}
\end{linenomath}
Here $\alpha_S$ and $\rho_S$ are absorption and Faraday mixing coefficients, respectively, in the Stokes basis.

We use a Cartesian coordinate system in the calculation of the transfer coefficients.  We set $\hat{z}$ parallel to the magnetic field $\mathbf{B}$.  The observer angle $\theta$ is the angle between $\mathbf{B}$ and the photon wavevector $\mathbf{k}$ which we choose to lie in the $x-z$ plane.  Note that we have chosen a coordinate system (in the plasma rest frame) such that all Stokes U coefficients are zero.  

Other recent work, such as \citet{huang_faraday_2011} and \citet{dexter_public_2016}, define $\mathbf{k}$ to lie in the $y-z$ plane.  This difference in convention requires one to change the sign of all Stokes Q and U coefficients when converting between our coefficients and theirs.  The signs of the Stokes I and V coefficients are the same in the two coordinate systems.  

We use cgs-Gaussian  units throughout.  The frequency of the photon is $\nu = c k/(2\pi)$, where $k$ is the magnitude of the wavevector.  We are often interested in coefficients' dependence on the ratio of the frequency to the cyclotron frequency, $\nu/\nu_c$, where $\nu_c = eB/(2\pi m_ec) = 2.8\times 10^7 B$.  Here $B$ is the magnitude of the magnetic field, $e$ is the elementary charge, and $m_e$ is the electron mass.
% Capitalize Gauss?
% use standard definitions

For an isotropic electron distribution, the  transfer coefficients are dependent solely on the distribution of electron Lorentz factors.\footnote{We do not consider anisotropic distribution functions. See \cite{Fleishman_2003} for a discussion} \citet{Melrose_1991} provide a procedure for calculating $j_S$ and $\alpha_S$ from the distribution function.  This procedure is implemented by \citet{leung_numerical_2011} in the \harmony{} code for a relativistic thermal distribution.  \citet{pandya_polarized_2016}, henceforth P16, introduced a simplified code, \symphony{}, which improves the accuracy and speed of the numerical integration.  In addition to the thermal distribution \symphony{} also provides coefficients for power-law and \kdf{} distributions.

Absorption and Faraday mixing coefficients are linearly related to components of the susceptibility tensor $\chi_{ij}$ (eqs. \ref{eqn:alphaSusceptibility}-\ref{eqn:rhoSusceptibility}).  \citet{huang_faraday_2011} describe a procedure and provide a Mathematica script to evaluate components of the susceptibility tensor for both a thermal and a power-law electron distribution function.  \citet{pandya_numerical_2018}, henceforth P18, extend the \symphony{} code to compute mixing and absorption coefficients via the susceptibility tensor method for the thermal, power-law, and \kdf{} distributions. 

The first goal of this paper is to review, update, and correct prior work. For example, we modify the sign of some transfer coefficient fits presented in P16 and P18 so that they are consistent with IEEE/IAU convention.  Second, we modify \symphony's method for evaluating Faraday conversion coefficients to permit efficient evaluation over a larger range of parameters.  Third, we provide new fitting formulae for Faraday mixing coefficients for \kdf{} distributions.

The plan of the paper is as follows.  Section 2 summarizes the mathematical construction of the integrals used to evaluate transfer coefficients for both the radiative transfer and the susceptibility tensor methods.  It also describes the distribution functions for which we compute transfer coefficients: thermal (Maxwell-Juettner) distributions; power-law distributions; and \kdf{} distributions.  Section 3 describes the numerical integration methods implemented in \symphony.  Section 4 presents updated fitting formulae for absorptivity and emissivity coefficients as well as new fitting formulae for Faraday mixing coefficients for \kdf{} distributions.

\section{Mathematical Construction}

\subsection{Radiative Transfer}

P16 use the radiative transfer equation in the Stokes basis to solve for polarized emissivity and absorption coefficients.  As in \citet{leung_numerical_2011} and P16, the polarized emissivity coefficients are given by

\begin{equation}
\label{eqn:emissivityVector}
    j_S = \begin{pmatrix} j_I \\ j_Q \\ j_U \\ j_V \end{pmatrix} = \frac{2\pi e^2\nu^2}{c}\int d^3p\,\,\,(f)\,\,\sum_{n = 1}^{\infty}\delta(y_n)K_S,
\end{equation}
and the absorptivity coefficients for an isotropic electron distribution function are given by
\begin{equation}
\label{eqn:absorptivityVector}
    \alpha_S = \begin{pmatrix} \alpha_I \\ \alpha_Q \\ \alpha_U \\ \alpha_V \end{pmatrix} = -\frac{\pi e^2}{m_ec}\int d^3p \,\,\left(\frac{\partial f}{\partial \gamma}\right)\,\, \sum_{n = 1}^{\infty}\delta(y_n)K_S,
\end{equation}
where $f$ is a normalized electron distribution function, $\gamma$ is the electron Lorentz factor, and $\delta$ is the Dirac delta function.  Eqs. \ref{eqn:thermal_f}-\ref{eqn:kappa_f} give $f$ for the thermal, power-law, and kappa distributions.  Here, $K_S$ depends on the Stokes parameter.  Complete definitions of $K_S$ and $y_n$ are given in \textcolor{black}{Appendix \ref{Appendix_A}}.

\symphony{} calculates emission and absorption coefficients by numerically evaluating the three-dimensional integral over momentum space and summing over harmonics, $n$.  The numerical methods used in calculating these transfer coefficients are summarized in section \ref{Radiative_Transfer_Numerical} of this paper. P16 provide a full description of the method.

\subsection{The Susceptibility Tensor}

We calculate the plasma susceptibility 3-tensor $\chi_{ij}$ in the $x$, $y$, $z$ basis, where $z$ is aligned with ${\mathbf B}$.  \textcolor{black}{A full statement of the susceptibility tensor is given in Appendix \ref{Appendix_B}}.  The definition of the susceptibility tensor and its relation to the transfer coefficients is derived in Section 2, Section 3, and Appendix A.1. of P18.  Absorption and Faraday mixing coefficients in the Stokes basis are related to the susceptibility tensor components by: 
\begin{equation}
\label{eqn:alphaSusceptibility}
    \alpha_S = \frac{\pi\nu}{c} \begin{cases}
     \text{Im}(\cos^2(\theta)\chi_{xx}-2\cos(\theta)\sin(\theta)\chi_{xz}+\sin^2(\theta)\chi_{zz}+\chi_{yy}), & \text{(Stokes I)}\\
     \text{Im}(\cos^2(\theta)\chi_{xx}-2\cos(\theta)\sin(\theta)\chi_{xz}+\sin^2(\theta)\chi_{zz}-\chi_{yy}), & \text{(Stokes Q)}\\
     \text{Im}(\cos(\theta)\chi_{yx}-\sin(\theta)\chi_{yz}+\cos(\theta)\chi_{xy}-\sin(\theta)\chi_{zy}) = 0,  & \text{(Stokes U)}\\
     \text{Re}(\cos(\theta)\chi_{xy}-\sin(\theta)\chi_{zy}-\cos(\theta)\chi_{yx}+\sin(\theta)\chi_{yz}), & \text{(Stokes V)}
    \end{cases}
\end{equation}
and
\begin{equation}
\label{eqn:rhoSusceptibility}
    \rho_S = \frac{\pi\nu}{c} \begin{cases}
      \text{Re}(\chi_{yy} - \cos^2(\theta)\chi_{xx} + 2\cos(\theta)\sin(\theta)\chi_{xz} - \sin^2(\theta)\chi_{zz}), & \text{(Stokes Q)}\\
      \text{Re}(\cos(\theta)\chi_{yx}-\sin(\theta)\chi_{yz}+\cos(\theta)\chi_{xy}-\sin(\theta)\chi_{zy}) = 0, & \text{(Stokes U)}\\
      \text{Im}(\cos(\theta)\chi_{xy}-\sin(\theta)\chi_{zy}-\cos(\theta)\chi_{yx}+\sin(\theta)\chi_{yz}), & \text{(Stokes V)}.
    \end{cases}
\end{equation}  
Here $\theta$ is the angle between the magnetic field ($z$-axis) and the wavevector, $\mathbf{k}$, which lies in the $x$-$z$ plane.  

The Stokes basis in the plasma frame is constructed so that projection of $\mathbf{B}$ in the polarization plane is along the $U>0$ axis, which sets $j_U$ to 0. As mentioned in P18, applying the Onsager relations to the susceptibility tensor (by requiring a time-reversal invariance of the microscopic dynamics) implies that $\chi_{xy}=-\chi_{yx}$, $\chi_{zy}=- \chi_{yz}$ and $\chi_{xz}=\chi_{xz}$. This relation implies $\alpha_U$ and $\rho_U$ are 0 as shown in Eq. \ref{eqn:alphaSusceptibility} and \ref{eqn:rhoSusceptibility}.  The Onsager relations may also be used to show that $\alpha_I$, $\alpha_Q$, and $\rho_Q$ are symmetric and $\alpha_V$ and $\rho_V$ are antisymmetric under a sign change of either the magnetic field or the particle's charge.  Similarly, it can be shown from the definition of $K_S$, given in \textcolor{black}{equation \ref{eqn:K_S}}, that $j_I$ and $j_Q$ are symmetric and $j_V$ is antisymmetric under one of these sign changes.

The susceptibility tensor components are evaluated in terms of a four-dimensional integral: three momentum space coordinates and a time coordinate $\tau$ that describes the unperturbed history of the electron orbit.  These integrals are performed over the derivative with respect to $\gamma$ of the scaled electron distribution function, $d\tilde{f}/d\gamma$.  Eqs. \ref{eqn:df_dgamma_thermal}-\ref{eqn:df_dgamma_kappa} provide $d\tilde{f}/d\gamma$ for the thermal, power-law, and kappa distributions.  P18 evaluate two of the momentum-space integrals analytically.  The remaining integrals are over $\gamma$ and $\tau$.  Section \ref{Suscept_Tensor_Method} summarizes the methods from P18 that \symphony{} uses to evaluate these integrals and details a new method for performing this integration for $\rho_Q$.

\subsection{Electron Distributions}

In order to evaluate transfer coefficients it is necessary to integrate over the distribution of electrons in three-dimensional momentum space:
\begin{equation}
    f \equiv \frac{dn_e}{d^3p} = \frac{1}{m_e^3c^3\gamma^2\beta}\frac{dn_e}{d\gamma d\cos\xi d\phi}.
\end{equation}
Here $\gamma$ is the electron Lorentz factor, $\xi$ is the pitch angle, $\beta = v/c$, where $v$ is the electron velocity, and $\phi$ is the gyrophase.  As in P16 and P18, we consider the relativistic thermal, the isotropic power-law, and the isotropic \kdf{} electron distributions.  

The thermal distribution is 
\begin{equation}
\label{eqn:thermal_f}
    \frac{dn_e}{d\gamma d\cos\xi d\phi} = \frac{n_e}{4\pi \Theta_e}\frac{\gamma(\gamma^2-1)^{1/2}}{K_2(1/\Theta_e)}\exp\left({-\frac{\gamma}{\Theta_e}}\right), \indent \text{(thermal)}
\end{equation}
where $\Theta_e \equiv k_B T/m_e c^2$ is the dimensionless electron temperature, $n_e$ is the number density of electrons, and $K_2$ is a modified Bessel function of the second kind.  

The power-law distribution is 
\begin{equation}
    \frac{dn_e}{d\gamma d\cos\xi d\phi} = \frac{n_e(p-1)}{4\pi (\gamma_{min}^{1-p} - \gamma_{max}^{1-p})}\gamma^{-p} \indent \text{for } \gamma_{min}\leq\gamma\leq\gamma_{max}, \indent \text{(power-law)}
\end{equation}
where $p$ is the power-law index and $\gamma_{max}$ and $\gamma_{min}$ are the maximum and minimum Lorentz factors.  

The \kdf{} distribution is 
\begin{equation}
\label{eqn:kappa_f}
    \frac{dn_e}{d\gamma d\cos\xi d\phi} =  \frac{n_eN_\kappa}{4\pi}\gamma (\gamma^2-1)^{1/2}\left(1+\frac{\gamma - 1}{\kappa w} \right)^{-(\kappa+1)}, \indent \text{(kappa)}
\end{equation}
where $\kappa$ is related to the high-energy power-law index, $w$ describes the width of the distribution, and $N_\kappa$ is a normalization factor that is evaluated numerically in \symphony{} to ensure that the integral of the distribution function over gamma is unity.

The absorptivities and rotativities depend on the derivatives $d\tilde{f}/d\gamma$, where $\tilde{f} \equiv m^3c^3f/n_e$. These derivatives are:
\begin{equation}
\label{eqn:df_dgamma_thermal}
    \frac{d\tilde{f}}{d\gamma} = -\frac{\exp{(-\gamma/\Theta_e)}}{4\pi \Theta_e^2K_2(1/\Theta_e)}, \indent \text{(thermal)}
\end{equation}

\begin{equation}
    \frac{d\tilde{f}}{d\gamma} = -\frac{(p-1)(-1+2\gamma^2+p(\gamma^2-1))}{4\pi (\gamma_{min}^{1-p}- \gamma_{max}^{1-p})\beta(\gamma^2-1)}\gamma^{-3-p}, \indent \text{(power-law)}
\end{equation}
and
\begin{equation}
\label{eqn:df_dgamma_kappa}
    \frac{d\tilde{f}}{d\gamma} = -\frac{N_\kappa(1+\kappa)}{4\pi\kappa w}\left(1+\frac{\gamma - 1}{\kappa w}\right)^{-2-\kappa}. \indent \text{(kappa)}
\end{equation}
Notice that this corrects a typographical error in P18's $d\tilde{f}/d\gamma$ for the power-law distribution.

\section{Numerical Methods} 

\subsection{Integration of Radiative Transfer Equations}
\label{Radiative_Transfer_Numerical}

Here we briefly summarize the numerical scheme used by \symphony{} to compute emission and absorption coefficients from equations \ref{eqn:emissivityVector} and \ref{eqn:absorptivityVector}.  More detail is provided in P16.  \symphony{} is based on \harmony{} (\citet{leung_numerical_2011}), with a simpler code organization and integration technique improvements that permit accurate evaluation of coefficients for Stokes V, and extension to larger $\nu/\nu_c$.

As in \citet{leung_numerical_2011}, P16 integrate equations \ref{eqn:emissivityVector} and \ref{eqn:absorptivityVector} over $\cos\xi$ and makes the substitution, $\cos\xi = (\nu-n\nu_c/\gamma)/(\nu\beta\cos\theta)$.  This reduces the integral and sum needed to compute emissivities and absorptivities to the form
\begin{equation}
    \int_{\gamma_-}^{\gamma_+}d\gamma\sum^{\infty}_{n=n_-}I(n,\gamma),
\end{equation}
where $I$ is the integrand which depends on the distribution function, the Stokes parameter, $\gamma_\pm = (r \pm |\cos\theta|(r^2-\sin^2\theta)^{1/2})/\sin^2\theta$, $r = n\nu_c/\nu$, and $n_- = (\nu|\sin\theta|)/\nu_c$.  Here $n_-$ is rounded up to become an integer. 

\symphony{} first calculates the $\gamma$ integral using the Quasi-Adaptive Gaussian quadrature routines, \texttt{QAG} and \texttt{QAGIU} (from GNU Science Library; GSL).  The $\gamma$ integration range is chosen based on accurate estimates for the location and width of the integrand's peak in $\gamma$ space.  This ensures that the quadrature captures the peak at large $\nu/\nu_c$, where  the integrand is sharply peaked. \symphony{} then directly computes the first 30 terms of the $n$ summation and approximates the remaining terms $(n \geq n_-+30)$ as an integral.  For Stokes V, the $\gamma$ integral is split into a positive part and a negative part with slightly different absolute areas on either side of the zero at $\gamma_0 = n\nu_c/(\nu\sin^2\theta)$.  The positive and negative contributions are summed over $n$ separately and then combined to avoid numerical error due to  cancellation. 

\subsection{Susceptibility Tensor Method} \label{Suscept_Tensor_Method}

To calculate absorption and rotation coefficients from components of the susceptibility tensor ($\chi_{ij}$) we numerically evaluate the integrals over $\tau$ and $\gamma$.  These integrals have the form
\begin{equation}
    \chi_{ij} = \int_0^{\infty}d\tau\int_1^\infty d\gamma \frac{d\tilde{f}}{d\gamma}  I_{ij}(\gamma, \tau, \nu/\nu_c, \mathbf{k}),
\end{equation}
where $I_{ij}$ is the integrand which depends on the component of the susceptibility tensor being calculated.  \textcolor{black}{Appendix \ref{Appendix_B}} provides a complete statement of these two-dimensional integrals \textcolor{black}{as presented in P18}.  Following P18 and integrating over $\tau$ first, the $\gamma$ integrand reduces to a smooth, well-behaved function.

For $\alpha_I$, $\alpha_V$, $\alpha_Q$, and $\rho_V$ we find the relevant components of the susceptibility tensor individually before combining them to obtain the transfer coefficients as in equations \ref{eqn:alphaSusceptibility} and \ref{eqn:rhoSusceptibility}. For $\rho_Q$, however, the combined susceptibility tensor components nearly cancel at large values of $\nu/\nu_c$, so small errors in individual components lead to large fractional errors in $\rho_Q$.  We minimize the fractional error by combining the integrands of the susceptibility tensor terms {\em before} integrating.

The tau integrand for $\rho_Q$ ($\equiv dK_{\rho_Q}$) oscillates rapidly and the integral is slow to converge.  We are able to numerically evaluate the integral, however, by using an  approximate form at large $\tau$ and the fact that the dominant contribution to the integral is at small $\tau$.  

Following \textcolor{black}{Appendix \ref{Appendix_B}}, the tau integrand oscillates at three distinct frequencies : $\omega_+(\tau)$,  $\omega_-(\tau)$, and $\omega_{env} = \omega_c/\omega$.  Here,
\begin{equation}
    \omega_{\pm}(\tau) = \gamma \pm \frac{A(\tau)}{\tau},
\end{equation}
where
\begin{equation}
    A(\tau) = \sqrt{\alpha^2+\delta^2},
\end{equation}
\begin{equation}
    \alpha = \gamma\beta\cos(\theta)\tau,
\end{equation}
and
\begin{equation}
    \delta = \frac{2\gamma\beta\omega\sin(\theta)}{\omega_c}\sin\left(\frac{\omega_c}{2\omega}\tau\right).
\end{equation}
The integral's dependence on $\omega_+$ and $\omega_-$ comes from the multiplication of sinusoidal factors with phases $A(\tau)$ and $\gamma\tau$ present in \textcolor{black}{equations \ref{eqn:Analytic_I_1_0}-\ref{eqn:Analytic_I_3_0} and \ref{eqn:Kernel_Suscept} of Appendix \ref{Appendix_B}}, respectively.  The integrand is multiplied by an envelope function that has frequency $\omega_{env}$ and decays like $\tau^{-1}$.  Integrating in steps of approximately $2\pi / \omega_+$ eliminates the dependence on $\omega_+$.  When $\tau \gtrsim \gamma\omega/\omega_c$, the integral can then be modelled as
\begin{equation}
    K = \int_0^\tau dK_{\rho_Q} \approx C + D\frac{\sin(\omega_{-}(\tau)\tau)}{\tau},
\end{equation}
where $C$ is the asymptotic value of the integral as $\tau \to \infty$ that we wish to evaluate and $D$ is an unimportant constant related to the amplitude of the envelope function.  When $\tau \gtrsim \gamma\omega/\omega_c$ and $\tau \gg 1$, the relative decay in the integral and the relative change in $\omega_-$ are small over a short interval in $\tau$.  We can therefore approximate the integral's form over a short interval in $\tau$ as, 
\begin{equation}
    K = C + E\sin(\omega_{-}\tau),
\end{equation}
where $E$ is another unimportant constant which has absorbed the approximately constant factor of $1/\tau$.  We then may calculate the asymptotic value of the integral through numerical evaluations of the integral and the second derivative of the integral using
\begin{equation}
    K'' = -\omega_{-}^2E\sin(\omega_{-}\tau),
\end{equation}
where each prime denotes a derivative with respect to $\tau$. Then,
\begin{equation}
    C = K + \frac{K''}{\omega_{-}^2}.
\end{equation}
We evaluate $K$, $K''$, and $\omega_-$ numerically using five equally spaced (by steps of $\Delta\tau = 2\pi / \omega_+$) evaluations of $K$ ($K_{-2}$-$K_{2}$):
\begin{equation}
    K'' \approx \frac{K_{1}-2K_{0}+K_{-1}}{\Delta\tau^2},
\end{equation}
and
\begin{equation}
    \omega_-^2 = -\frac{K'''}{K'} \approx \frac{K''_{-1}-K''_{1}}{K_{1}-K_{-1}} \approx \frac{K_{-2}-2K_{-1}+2K_{1}-K_{2}}{\Delta\tau^2(K_{1}-K_{-1})}.
\end{equation}
Then
\begin{equation}
    C \approx \frac{K_{1}^2-K_{-1}^2+(K_{0}(K_{-2}-K_{2}))}{K_{-2}-2K_{-1}+2K_{1}-K_{2}}.
\end{equation}

\begin{figure}
\centering 
\subfloat[]{%
  \includegraphics[width=0.48\columnwidth]{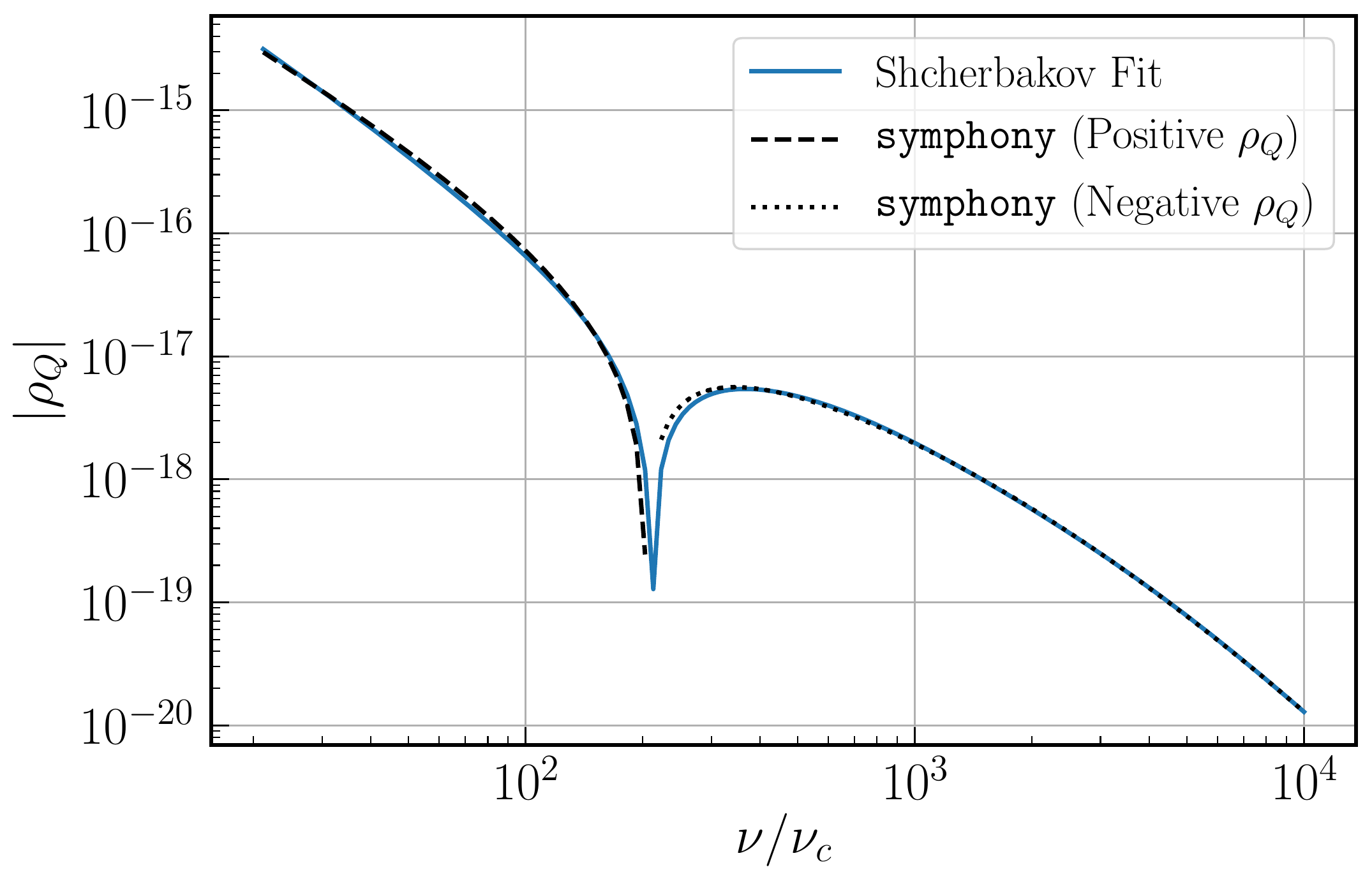}%
  \label{fig:1a}%
}\qquad
\subfloat[]{%
  \includegraphics[width=0.48\columnwidth]{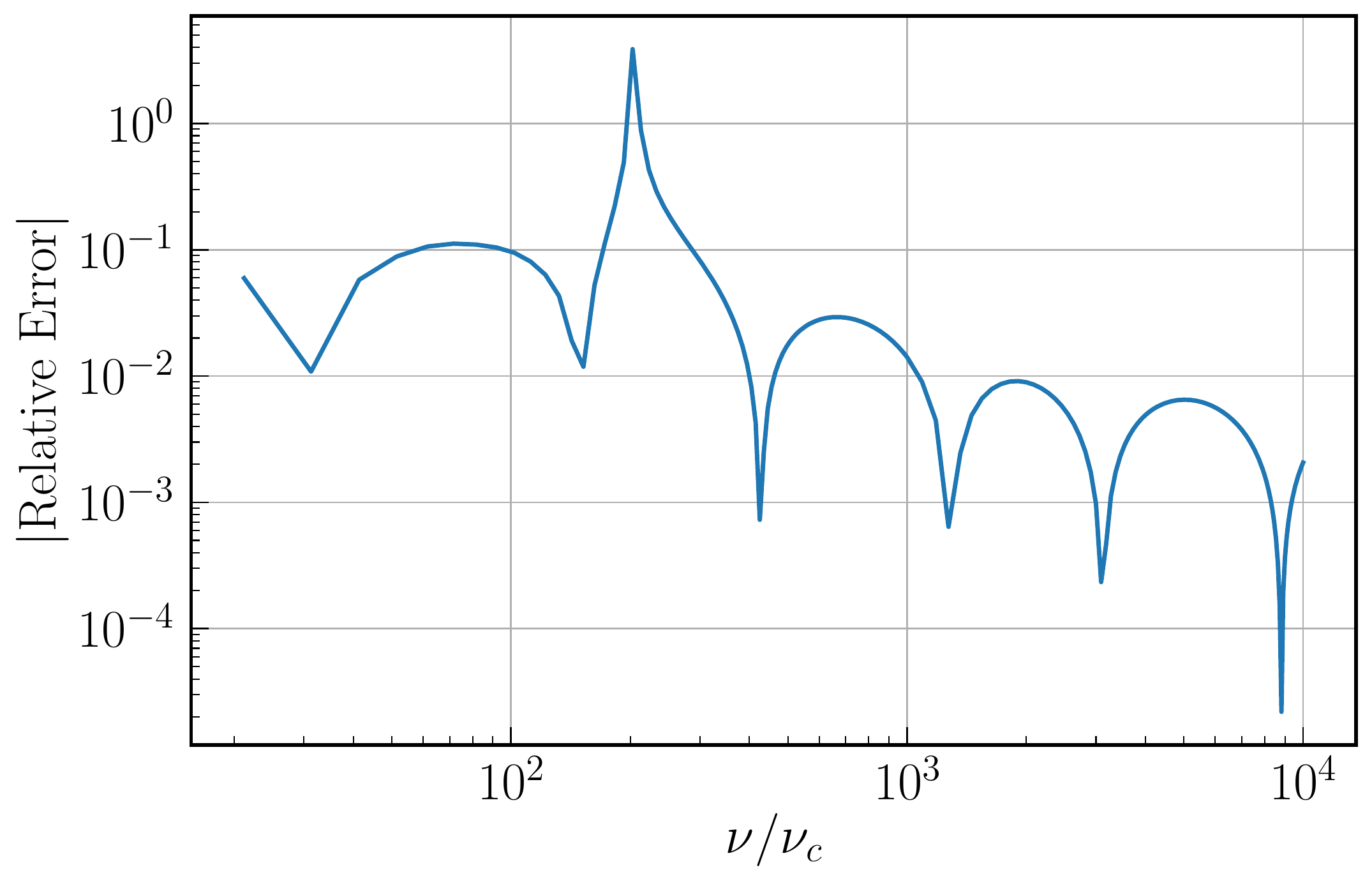}%
  \label{fig:1b}%
}
\caption{Comparison of \symphony{}'s numerically evaluated $\rho_Q$ to the thermal $\rho_Q$ fit presented in \citet{shcherbakov_propagation_2008}. Panel (a) is a log-log scale plot of $|\rho_Q|$ versus $\nu/\nu_c$ for a thermal ($\Theta_e = 10$, $\theta = 60^{\circ}$) distribution.  The Shcherbakov fit is represented by the solid line and \symphony{}'s numerical evaluations are represented by the marks.  The sign change in $\rho_Q$ is represented in panel (a) by the change from dashed to dotted linestyles.  Relative error between the two versus $\nu/\nu_c$ on a log scale is shown in panel (b).}
\label{fig:1}
\end{figure}

Fig. \ref{fig:1} compares Faraday conversion coefficients calculated using this method to the thermal fit for $\rho_Q$ presented in \citet{shcherbakov_propagation_2008}.  Throughout this paper we define relative error as
\begin{equation}
    \text{Relative Error} = \frac{\text{Fitted Value}}{\text{Numerically Calculated Value}} - 1.
\end{equation}

%\begin{figure}
%     \begin{subfigure}[h!]{0.49\textwidth}
%         \includegraphics[width=\textwidth]{rhoQThermal.pdf}
%         \label{fig:1a}
%         \centering
%         \caption{}
%     \end{subfigure}
%     \begin{subfigure}[h!]{0.49\textwidth}
%         \includegraphics[width=\textwidth]{rhoQThermalError.pdf}
%         \label{fig:1b}
%         \caption{}
%     \end{subfigure}
%        \caption{Comparison of \symphony{}'s numerically evaluated $\rho_Q$ to the thermal $\rho_Q$ fit presented in \citet{shcherbakov_propagation_2008}. Panel (a) is a log-log scale plot of $|\rho_Q|$ versus $\nu/\nu_c$ for a thermal ($\Theta_e = 10$, $\theta = 60^{\circ}$) distribution.  The Shcherbakov fit is represented by the solid line and \symphony{}'s numerical evaluations are represented by the marks.  The sign change in $\rho_Q$ is represented in panel (a) by the change from dashed to dotted linestyles.  Relative error between the two versus $\nu/\nu_c$ on a log scale is shown in panel (b).}
%\label{fig:1}
%\end{figure}

\section{Updated Fits for Transfer Coefficients}

%% Include all fits for Thermal, Power Law, and Kappa Distributions
%% Reference and mention corrections to earlier Pandya papers

Here we provide fitting formulae for the complete set of absorptivity and emissivity coefficients in the Stokes basis.  Unless stated otherwise the fitting formulae are identical to those in P16.  

The emissivity has the general form 
\begin{equation}
    j_S = \frac{n_e e^2 \nu_c}{c}J_S\left(\frac{\nu}{\nu_c},\theta\right),
\end{equation}
and the absorptivity has the general form
\begin{equation}
    \alpha_S = \frac{n_e e^2 }{\nu m_e c}A_S\left(\frac{\nu}{\nu_c},\theta\right).
\end{equation}
Here $J_S$ and $A_S$ are dimensionless functions of the distribution specific parameters: $\Theta_e$ for thermal, $p$, $\gamma_{min}$, and $\gamma_{max}$ for power-law, and $w$ and $\kappa$ for \kdf{} distributions.

\subsection{Thermal Distribution}

We have replaced our $J_V$ fit for a thermal distribution with the fit originally presented in \citet{dexter_public_2016} (eqs. A14 and A20 in Appendix A).  We approximate this fit with rational coefficients and recast it in our notation.  The dimensionless emissivity is 
\begin{equation}
\label{eqn:J_S_thermal}
J_S = \exp{(-X^{1/3})}
    \times \begin{cases}
    \frac{\sqrt{2}\pi}{27}\sin{\theta}(X^{1/2}+2^{11/12}X^{1/6})^2, & \text{(Stokes I)}\\
     -\frac{\sqrt{2}\pi}{27}\sin{\theta} \left(X^{1/2}+\frac{7\Theta_e^{24/25}+35}{10\Theta_e^{24/25}+75}2^{11/12}X^{1/6}\right)^2, & \text{(Stokes Q)}\\ 
     0, & \text{(Stokes U)}\\
     \frac{1}{\Theta_e}\cos\theta\left(\frac{\pi}{3}+\frac{\pi}{3}X^{1/3}+(\frac{2}{300})X^{1/2}+(\frac{2\pi}{19})X^{2/3}\right). & \text{(Stokes V)}
    \end{cases}
\end{equation}
Here $X = \nu/\nu_s$, where $\nu_s \equiv (2/9)\nu_c\sin{\theta}\Theta_e^2$.

For a thermal distribution we can use Kirchoff's law to obtain the absorptivity:
\begin{equation}
\label{eqn:Kirchoffs}
    j_S -\alpha_S B_\nu = 0,
\end{equation}
where $B_\nu \equiv (2h\nu^3/c^2)[\exp(h\nu/kT)-1]^{-1}$ is the Planck function.  Equation \ref{eqn:Kirchoffs} corrects an error in equation 25 of P16. The dimensionless absorptivity is then
%% Maybe put in terms of \Theta_e instead of T
\begin{equation}
\label{eqn:KirchoffsDimensionless}
    A_S = J_S\frac{m_ec^2\nu_c}{2h\nu^2}(e^{h\nu / (kT)} - 1).
\end{equation}
Equations \ref{eqn:Kirchoffs} and \ref{eqn:KirchoffsDimensionless} (eqs. 25 and 32 in P16) were incorrectly combined in equation 32 of P16.  Here it should be clear that equation \ref{eqn:Kirchoffs} applies to $j_S$ and $\alpha_S$, while equation \ref{eqn:KirchoffsDimensionless} applies to the dimensionless $J_S$ and $A_S$. 

\citet{shcherbakov_propagation_2008} provides fitting formulae for thermal rotativities that maintain accuracy across high frequencies ($X \gg 1$) and high temperatures ($\Theta_e \gtrsim 1$).  \citet{dexter_public_2016} modifies these expressions to maintain accuracy for smaller $\nu$.  These modified fits, in our notation \textcolor{black}{and sign convention}, are
\begin{equation}
    \rho_Q = \textcolor{black}{-}\frac{n_ee^2\nu_c^2\sin^2\theta}{mc\nu^3} f_m(X) \left[\frac{K_1(\Theta_e^{-1})}{K_2(\Theta_e^{-1})}+6\Theta_e\right],
\end{equation}
where

\begin{equation}
    f_m(X) = f_0(X) + \left[0.011\exp\left(-1.69X^{-1/2}\right) - 0.003135X^{4/3} \right]\left(\frac{1}{2}[1+\tanh(10\ln(0.6648X^{-1/2}))]\right),
\end{equation}
with
\begin{equation}
    f_0(X) = 2.011\exp\left(-19.78X^{-0.5175}\right) - \cos\left(39.89X^{-1/2}\right) \exp\left(-70.16X^{-0.6}\right) - 0.011\exp\left(-1.69X^{-1/2}\right),
\end{equation}
and
\begin{equation}
    \rho_V = \frac{2n_ee^2\nu_c}{mc\nu^2} \frac{K_0(\Theta_e^{-1}) - \Delta J_5(X)}{K_2(\Theta_e^{-1})} \cos\theta,
\end{equation}
where
\begin{equation}
    \Delta J_5(X) = 0.4379\ln(1+1.3414X^{-0.7515}),
\end{equation}
and $K_n$ is the modified Bessel function of the second kind and order n.  These expressions maintain accuracy for all $X$ where $\nu/\nu_c \gg 1$.

\subsection{Power-Law Distribution Fits}

The dimensionless emissivities are
\begin{linenomath}
\begin{align}
J_S &= \frac{3^{p/2}(p-1)\sin{\theta}}{2(p+1)(\gamma_{min}^{1-p}-\gamma_{max}^{1-p})} \nonumber \\ &\times \Gamma\left(\frac{3p-1}{12}\right)\Gamma\left(\frac{3p+19}{12}\right)\left(\frac{\nu}{\nu_c\sin\theta}\right)^{-(p-1)/2} \nonumber
    \\ &\times \begin{cases}
     1 & \text{(Stokes I)}\\
     -\frac{p+1}{p+7/3} & \text{(Stokes Q)}\\ 
     0 & \text{(Stokes U)}\\
     \frac{171}{250}\frac{p^{49/100}}{\tan\theta}\left(\frac{\nu}{3\nu_c \sin\theta}\right)^{-1/2}& \text{(Stokes V)}.
    \end{cases}
\end{align}
\end{linenomath}
Note that $J_V$ has changed sign compared to P16 so that it is now consistent with IEEE/IAU conventions.

Kirchoff's law cannot be used for non-thermal distributions.  P16 fit the dimensionless absorptivities with
\begin{linenomath}
\begin{align}
A_S &= \frac{3^{(p+1)/2}(p-1)}{4(\gamma_{min}^{1-p}-\gamma_{max}^{1-p})} \nonumber \\ &\times \Gamma\left(\frac{3p+2}{12}\right)\Gamma\left(\frac{3p+22}{12}\right)\left(\frac{\nu}{\nu_c\sin\theta}\right)^{-(p+2)/2} \nonumber
    \\ &\times \begin{cases}
     1 & \text{(Stokes I)}\\
     -(\frac{17}{500}p-\frac{43}{1250})^{43/500} & \text{(Stokes Q)}\\ 
     0 & \text{(Stokes U)}\\
     \left(\frac{71}{100}p+\frac{22}{625}\right)^{197/500}(\frac{31}{10}(\sin \theta)^{-48/25}-\frac{31}{10})^{64/125}\left(\frac{\nu}{\nu_c \sin \theta}\right)^{-1/2}\sign({\cos\theta}), & \text{(Stokes V)}
    \end{cases}
\end{align}
\end{linenomath}
\textcolor{black}{where $\sign(x)$ is the sign function which extracts the sign of its argument.} This expression corrects a typographical error in the argument of the first gamma function in P16, and we have changed the sign of $A_V$ to be consistent with the IEEE/IAU convention.  We have also introduced a factor of $\sign({\cos\theta})$ to make $J_V$ antisymmetric about $\theta = \pi/2$.

\begin{figure}
\centering 
\subfloat[]{%
  \includegraphics[width=0.48\columnwidth]{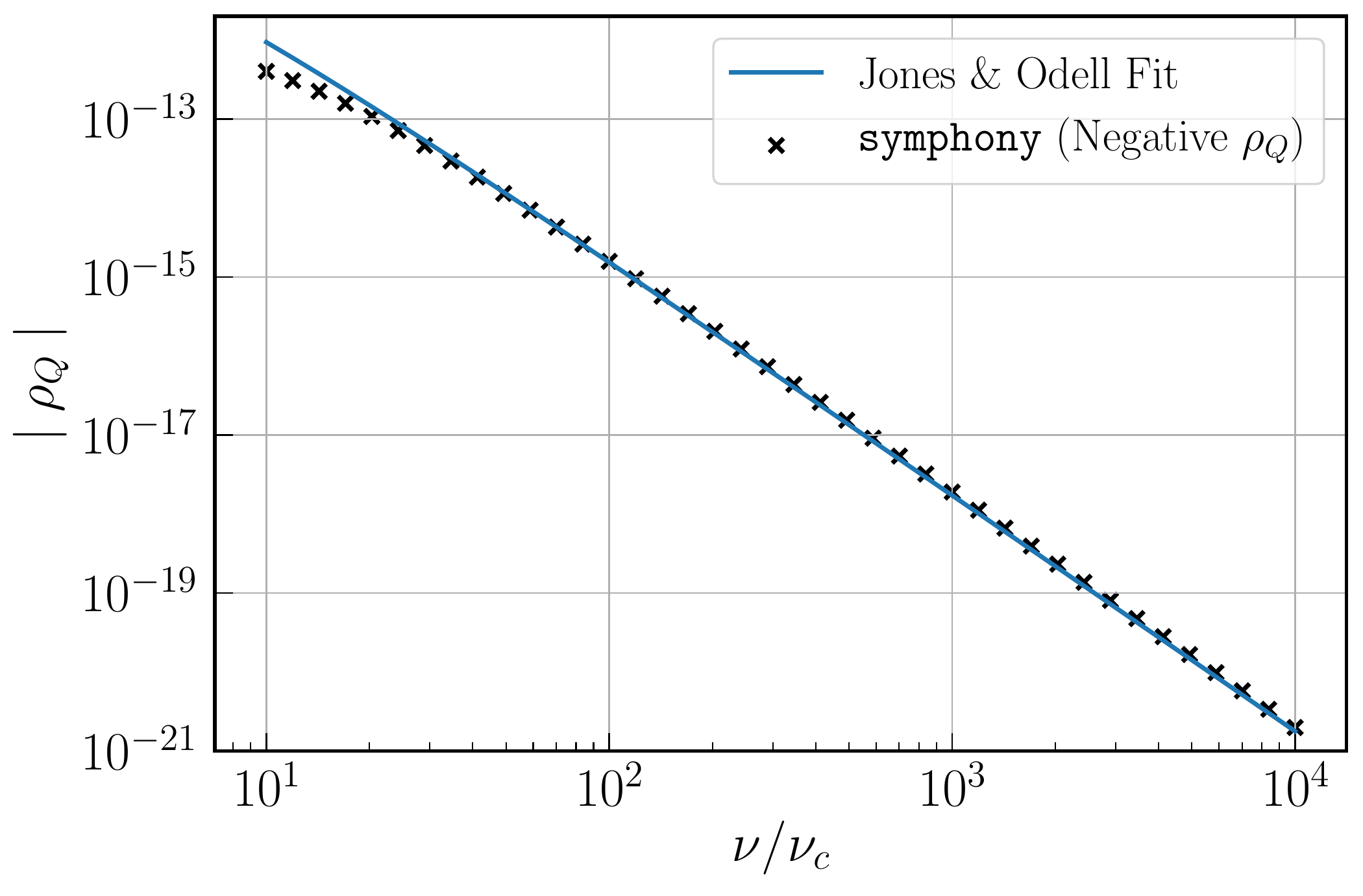}%
  \label{fig:3a}%
}\qquad
\subfloat[]{%
  \includegraphics[width=0.48\columnwidth]{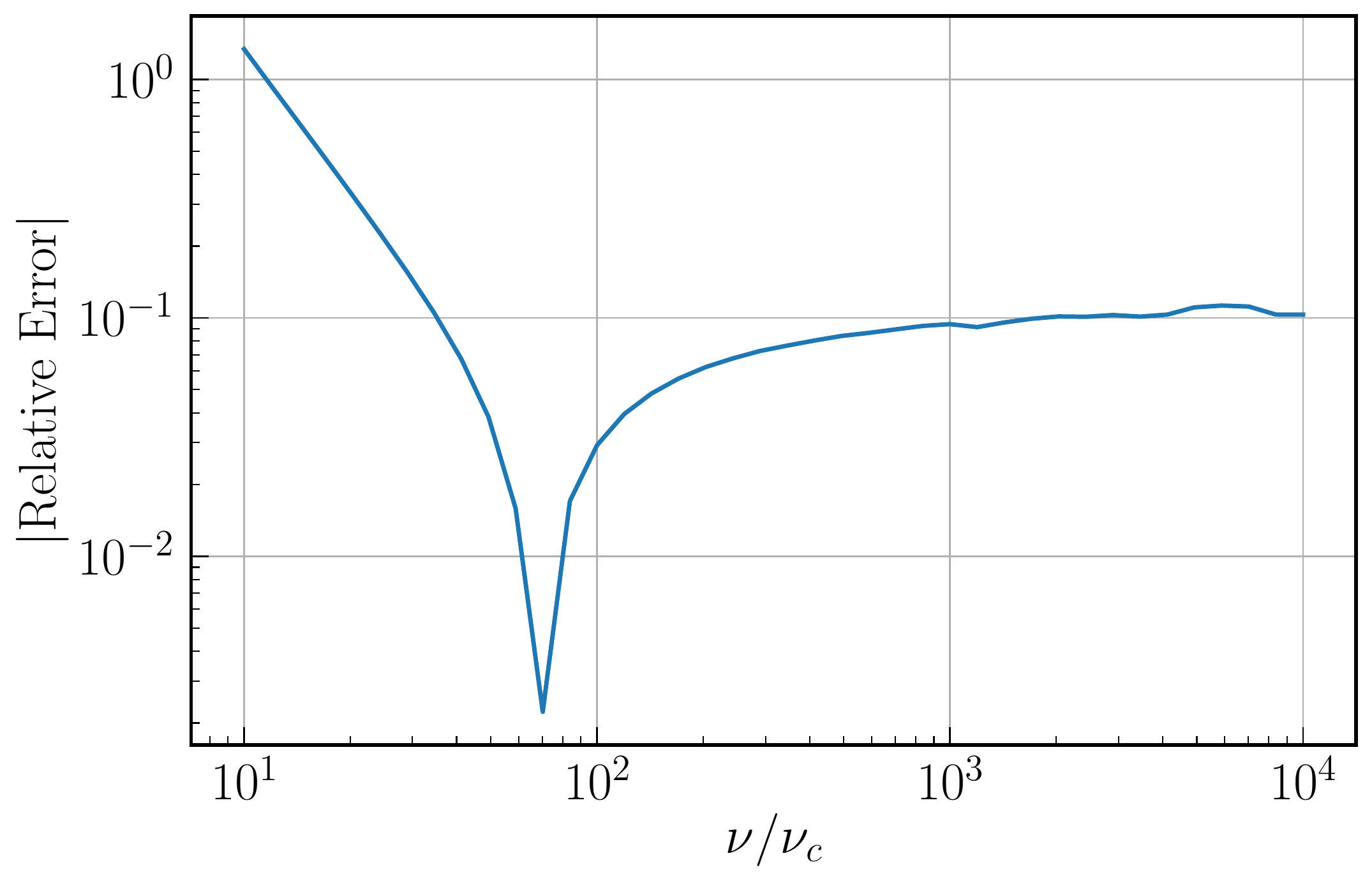}%
  \label{fig:3b}%
}
\caption{Comparison of \symphony{}'s numerically evaluated $\rho_Q$ to the power-law $\rho_Q$ fit presented in \citet{Jones_1977}.  For this range of parameters, $\rho_Q < 0$.  Panel (a) is a log-log scale plot of $|\rho_Q|$ versus $\nu/\nu_c$ for a power-law ($p = 3$, $\gamma_{min} = 2$, $\gamma_{max} = 1000$, $\theta = 60^{\circ}$) distribution.  The Jones $\&$ Odell fit is represented by the solid line and \symphony{}'s numerical evaluations are represented by the $x$ marks.  Relative error between the two versus $\nu/\nu_c$ on a log scale is shown in panel (b).}
\label{fig:PLCompare}
\end{figure}

\citet{Jones_1977} provide approximate rotativities for the power-law distribution.  Their fits, written in our notation, are
\begin{equation}
    \rho_Q = -\rho_\bot \left(\frac{\nu_c\sin\theta}{\nu}\right)^3 \frac{\gamma_{min}^{2-p}}{(p/2)-1} \left[1 - \left(\frac{2\nu_c\gamma_{min}^2\sin\theta}{3\nu}\right)^{p/2-1}\right],
    \label{eqn:JOrhoQ}
\end{equation}
\begin{equation}
    \rho_V = 2\rho_{\bot}\frac{p+2}{p+1} \left(\frac{\nu_c\sin\theta}{\nu}\right)^2 \gamma_{min}^{-(p+1)} \ln{(\gamma_{min})} \cot\theta,
\end{equation}
where
\begin{equation}
    \rho_\bot = \frac{n_ee^2}{mc\nu_c\sin\theta}(p-1) \left[\gamma_{min}^{1-p} - \gamma_{max}^{1-p}\right]^{-1}.
\end{equation}
These fits are relatively accurate for $\gamma_{min} \lesssim 10^2$ where $\nu/\nu_c \gg 1$.  A comparison of the $\rho_Q$ fit in Eq. \ref{eqn:JOrhoQ} to \symphony{}'s numerically evaluated $\rho_Q$ is shown in Fig. \ref{fig:PLCompare}.

%\begin{figure}
%     \begin{subfigure}[h!]{0.49\textwidth}
%         \includegraphics[width=\textwidth]{rhoQPL.pdf}
%         \label{fig:3a}
%         \centering
%         \caption{}
%     \end{subfigure}
%     \begin{subfigure}[h!]{0.49\textwidth}
%         \includegraphics[width=\textwidth]{rhoQPLError.pdf}
%         \label{fig:3b}
%         \caption{}
%     \end{subfigure}
%        \caption{Comparison of \symphony{}'s numerically evaluated $\rho_Q$ to the power-law $\rho_Q$ fit presented in \citet{Jones_1977}.  For this range of parameters, $\rho_Q < 0$.  Panel (a) is a log-log scale plot of $|\rho_Q|$ versus $\nu/\nu_c$ for a power-law ($p = 3$, $\gamma_{min} = 2$, $\gamma_{max} = 1000$, $\theta = 60^{\circ}$) distribution.  The Jones $\&$ Odell fit is represented by the solid line and \symphony{}'s numerical evaluations are represented by the $x$ marks.  Relative error between the two versus $\nu/\nu_c$ on a log scale is shown in panel (b).}
%        \label{fig:PLCompare}
%\end{figure}

\subsection{Kappa Distribution Fits}

P16 fit absorptivities and emissivities for kappa distributions by separately fitting the high-frequency and low-frequency limits and providing a bridging function between these limits.   In terms of $\nu_{\kappa} \equiv  \nu_c(w\kappa)^2\sin\theta$ and  $X_\kappa \equiv  \nu/\nu_\kappa$, the dimensionless emissivities in the low-frequency limit are
\begin{linenomath}
\begin{align}
J_{S,lo} &= X_\kappa^{1/3}\sin(\theta)\frac{4\pi\Gamma(\kappa-4/3)}{3^{7/3}\Gamma(\kappa-2)} \nonumber \\
    &\times \begin{cases}
    1, & \text{(Stokes I)}\\
    \frac{1}{2}, & \text{(Stokes Q)}\\ 
     0, & \text{(Stokes U)}\\
     \left(\frac{3}{4}\right)^2[(\sin \theta)^{-12/5}-1]^{12/25}\frac{\kappa^{-66/125}}{w}X_\kappa^{-7/20}. & \text{(Stokes V)}
    \end{cases}
\end{align}
\end{linenomath}

The dimensionless emissivities in the high-frequency limit are
\begin{linenomath}
\begin{align}
J_{S,hi} &= X_\kappa^{-(\kappa-2)/2}\sin(\theta)3^{(\kappa-1)/2}\frac{(\kappa-2)(\kappa-1)}{4}\Gamma\left(\frac{\kappa}{4}-\frac{1}{3}\right)\Gamma\left(\frac{\kappa}{4}+\frac{4}{3}\right) \nonumber \\
    &\times \begin{cases}
    1, & \text{(Stokes I)}\\
    \left[\left(\frac{4}{5}\right)^2+\frac{\kappa}{50}\right], & \text{(Stokes Q)}\\ 
     0, & \text{(Stokes U)}\\
     \left(\frac{7}{8}\right)^2[(\sin \theta)^{-5/2}-1]^{11/25}\frac{\kappa^{-11/25}}{w}X_\kappa^{-1/2}. & \text{(Stokes V)}
    \end{cases}
\end{align}
\end{linenomath}

The emissivity bridging function is
\begin{equation}
    J_S = \begin{cases}
    (J_{S,lo}^{-x}+J_{S,hi}^{-x})^{-1/x}, & \text{(Stokes I)} \\
    -(J_{S,lo}^{-x}+J_{S,hi}^{-x})^{-1/x}, & \text{(Stokes Q)} \\
    (J_{S,lo}^{-x}+J_{S,hi}^{-x})^{-1/x}\sign({\cos\theta}), & \text{(Stokes V)} \\
    \end{cases}
\end{equation}
where
\begin{linenomath}
\begin{align}
    \label{eqn:x_A_S}
    x = \begin{cases}
    3\kappa^{-3/2}, & \text{(Stokes I)}\\
    \frac{37}{10}\kappa^{-8/5}, & \text{(Stokes Q)}\\ 
    3\kappa^{-3/2}. & \text{(Stokes V)}
    \end{cases}
\end{align}
\end{linenomath}
Notice that we have made sign corrections to $J_Q$ and $J_V$ compared to equations 35-37 of P16.  We also multiply $J_V$ by an overall factor of $\sign({\cos\theta})$ to make it antisymmetric about $\theta = \pi/2$.  The expression for $x$ for Stokes V has also been updated.

The dimensionless absorptivities in the low-frequency limit are
\begin{linenomath}
\begin{align}
A_{S,lo} &= X_\kappa^{-2/3}3^{1/6}\frac{10}{41}\frac{(2\pi)}{(w\kappa)^{10/3-\kappa}}\frac{(\kappa-2)(\kappa-1)\kappa}{3\kappa-1} \nonumber \\
    &\times \Gamma\left(\frac{5}{3}\right) {}_{2}F_{1}\left(\kappa-\frac{1}{3}, \kappa+1, \kappa +\frac{2}{3}, -\kappa w\right) \nonumber \\
    &\times \begin{cases}
    1, & \text{(Stokes I)}\\
    \frac{25}{48}, & \text{(Stokes Q)}\\ 
     0, & \text{(Stokes U)}\\
    \frac{77}{100w}[(\sin \theta)^{-114/50}-1]^{223/500}X_\kappa^{-7/20} \kappa^{-7/10}. & \text{(Stokes V)}
    \end{cases}
\end{align}
\end{linenomath}

The dimensionless absorptivities in the high-frequency limit are
\begin{linenomath}
\begin{align}
A_{S,hi} &= X_\kappa^{-(1+\kappa)/2}\frac{\pi^{3/2}}{3}\frac{(\kappa-2)(\kappa-1)\kappa}{(w\kappa)^3}  \nonumber \\
    &\times \left(\frac{2\Gamma(2+\kappa/2)}{2+\kappa}-1\right)  \nonumber \\
    &\times \begin{cases}
    \left(\left(\frac{3}{\kappa}\right)^{19/4}+\frac{3}{5}\right), & \text{(Stokes I)}\\
    \left(21^2\kappa^{-(12/5)^2}+\frac{11}{20}\right), & \text{(Stokes Q)}\\ 
     0, & \text{(Stokes U)}\\
    \frac{143}{10}w^{-116/125}[(\sin \theta)^{-41/20}-1]^{1/2}\left\{13^2\kappa^{-8}+\frac{13}{2500}\kappa-\frac{263}{5000}+\frac{47}{200\kappa}\right\}X_\kappa^{-1/2}. & \text{(Stokes V)}
    \end{cases}
\end{align}
\end{linenomath}

The absorptivity bridging function is
\begin{equation}
    A_S = \begin{cases}
    (A_{S,lo}^{-x}+A_{S,hi}^{-x})^{-1/x}, & \text{(Stokes I)} \\
    -(A_{S,lo}^{-x}+A_{S,hi}^{-x})^{-1/x}, & \text{(Stokes Q)} \\
    (A_{S,lo}^{-x}+A_{S,hi}^{-x})^{-1/x}\sign({\cos\theta}), & \text{(Stokes V)} \\
    \end{cases}
\end{equation}
where
\begin{linenomath}
\begin{align}
    x = \begin{cases}
    \left(-\frac{7}{4}+\frac{8}{5}\kappa\right)^{-43/50}, & \text{(Stokes I)}\\
    \frac{7}{5}\kappa^{-23/20}, & \text{(Stokes Q)}\\ 
    \frac{61}{50}\kappa^{-142/125}+\frac{7}{1000}. & \text{(Stokes V)}
    \end{cases}
\end{align}
\end{linenomath}
We have again made sign corrections to the absorptivities for Stokes Q and Stokes V compared to equations 38-40 of P16.  We also multiply $A_V$ by a factor of $\sign({\cos\theta})$ to make it antisymmetric about $\theta = \pi/2$. The third term in curly braces for $A_{V,hi}$ has been changed from its original presentation to improve the fit's accuracy.

%%Andy: why was A_V changed?

We provide fitting formulae for $\rho_V$ and $\rho_Q$ for four \kdf{} distributions between $\kappa = 3.5$ and $\kappa = 5$.  The structure of these fits is based on Faraday mixing coefficient fits for a thermal distribution provided in \citet{shcherbakov_propagation_2008}. 
\begin{linenomath}
\begin{align}
\rho_Q &=  -\frac{n_ee^2\nu_c^2\sin^2\theta}{mc\nu^3}f(X_\kappa) \nonumber \label{eqn:rho_Q_kappa}\\
    & \times \begin{cases}
     17w-3\sqrt{w}+7\sqrt{w}\exp({-5w}), & (\kappa = 3.5)\\
     \frac{46}{3}w-\frac{5}{3}\sqrt{w}+\frac{17}{3}\sqrt{w}\exp({-5w}), & (\kappa = 4)\\ 
     14w-\frac{13}{8}\sqrt{w}+\frac{9}{2}\sqrt{w}\exp({-5w}), & (\kappa = 4.5)\\
     \frac{25}{2}w-\sqrt{w}+5\sqrt{w}\exp({-5w}), & (\kappa = 5)
    \end{cases}
\end{align}
\end{linenomath}
where
\begin{equation}
f(X_\kappa) =  
    \begin{cases}
     1 - \exp\left(-\frac{X^{0.84}}{30}\right) - \sin\left(\frac{X}{10}\right)\exp\left(-\frac{3X^{0.471}}{2}\right), & (\kappa = 3.5)\\
     1 - \exp\left(-\frac{X^{0.84}}{18}\right) - \sin\left(\frac{X}{6}\right)\exp\left(-\frac{7X^{0.5}}{4}\right), & (\kappa = 4)\\ 
     1 - \exp\left(-\frac{X^{0.84}}{12}\right) - \sin\left(\frac{X}{4}\right)\exp\left(-2X^{0.525}\right), & (\kappa = 4.5)\\
     1 - \exp\left(-\frac{X^{0.84}}{8}\right) - \sin\left(\frac{3X}{8}\right)\exp\left(-\frac{9X^{0.541}}{4}\right). & (\kappa = 5)
    \end{cases}
\end{equation}
%\begin{figure}[!h]
%     \begin{subfigure}[h!]{0.49\textwidth}
%         \includegraphics[width=.95\textwidth]{rhoQScatter.pdf}
%         \label{fig:2a}
%         \centering
%         \caption{}
%     \end{subfigure}
%     \begin{subfigure}[h!]{0.49\textwidth}
%         \includegraphics[width=.95\textwidth]{rhoQRelError.pdf}
%         \label{fig:2b}
%         \caption{}
%     \end{subfigure}
%     \begin{subfigure}[h!]{0.49\textwidth}
%         \includegraphics[width=.95\textwidth]{rhoVScatter.pdf}
%         \label{fig:2c}
%         \caption{}
%     \end{subfigure}
%        \begin{subfigure}[h!]{0.49\textwidth}
%         \includegraphics[width=.95\textwidth]{rhoVRelError.pdf}
%         \caption{}
%         \label{fig:2d}
%     \end{subfigure}
%        \caption{Comparison of our fits to \symphony{}'s numerically evaluated Faraday mixing coefficients. Panels (a) and (c) are log-log scale plots of $|\rho_Q|$ and $\rho_V$, respectively, versus $X_\kappa$ for various \kdf{} distributions.  The fitted values are shown by the solid line and  \symphony{}'s numerical evaluations are represented by the marks.  The sign change in $\rho_Q$ is represented in panel (a) by the change from "$+$" marks to "-" marks.  Relative error of the fits versus $X_\kappa$ on a log scale is shown in panels (b) and (d).  For these plots we set $w = 4$, $\theta = 60^{\circ}$, $B = 10$ Gauss, and varied $\nu$ to capture a range of $X_\kappa$.}
%        \label{fig:2}
%\end{figure}
\begin{figure}
\centering 
\subfloat[]{%
  \includegraphics[width=0.48\columnwidth]{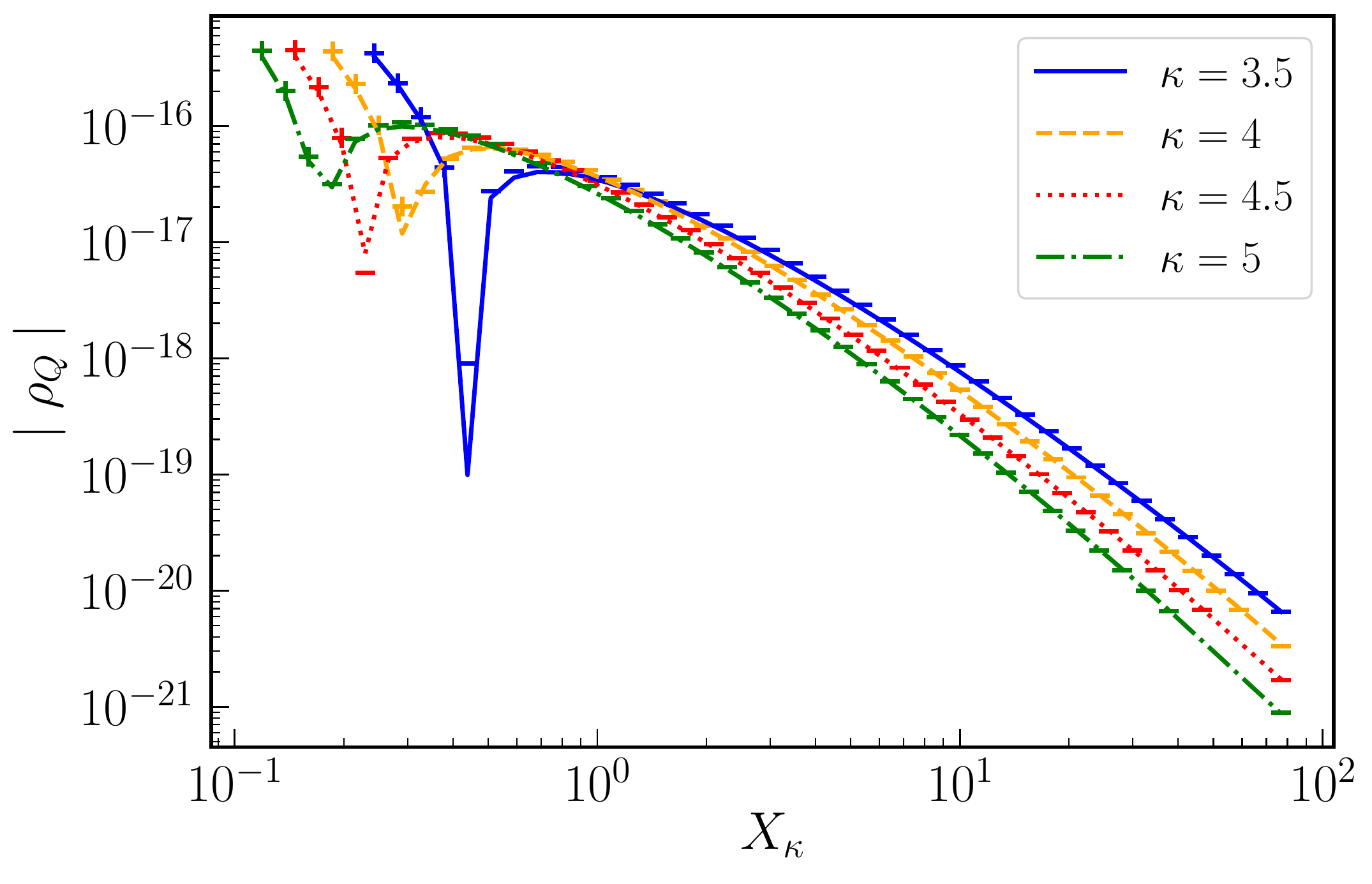}%
  \label{fig:2a}%
}\qquad
\subfloat[]{%
  \includegraphics[width=0.48\columnwidth]{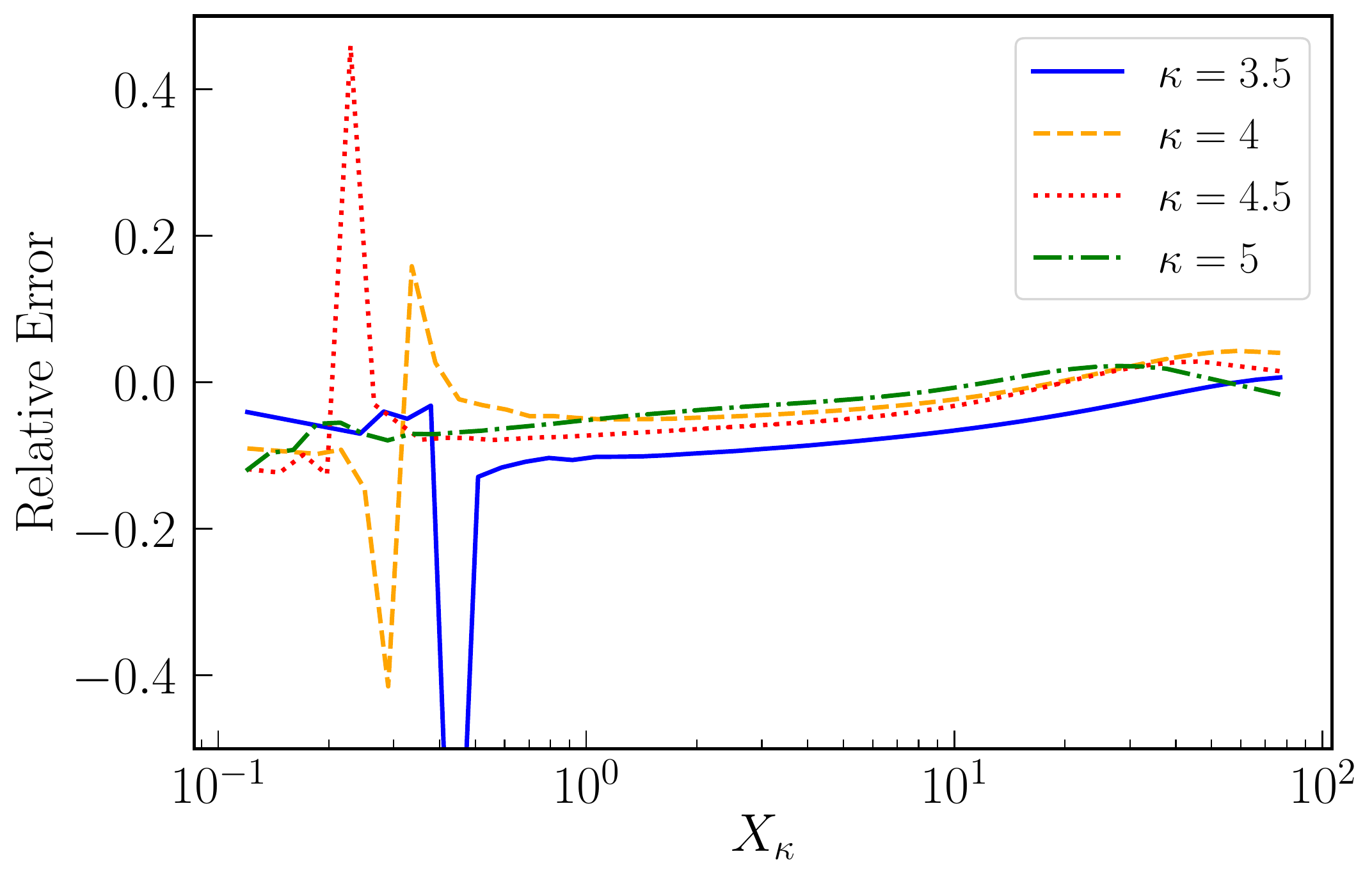}%
  \label{fig:2b}%
}\qquad
\subfloat[]{%
  \includegraphics[width=0.48\columnwidth]{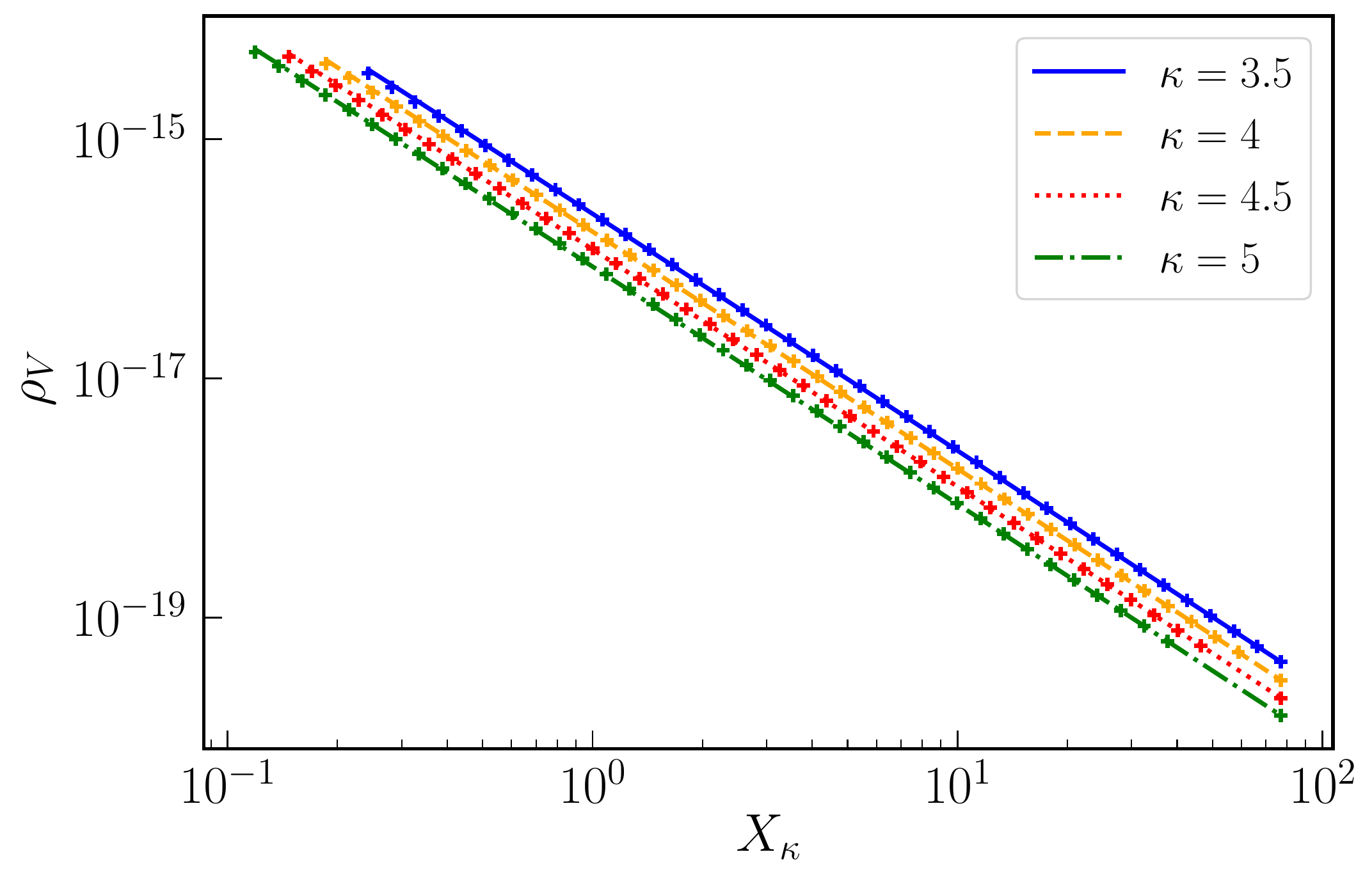}%
  \label{fig:2c}%
}\qquad
\subfloat[]{%
  \includegraphics[width=0.48\columnwidth]{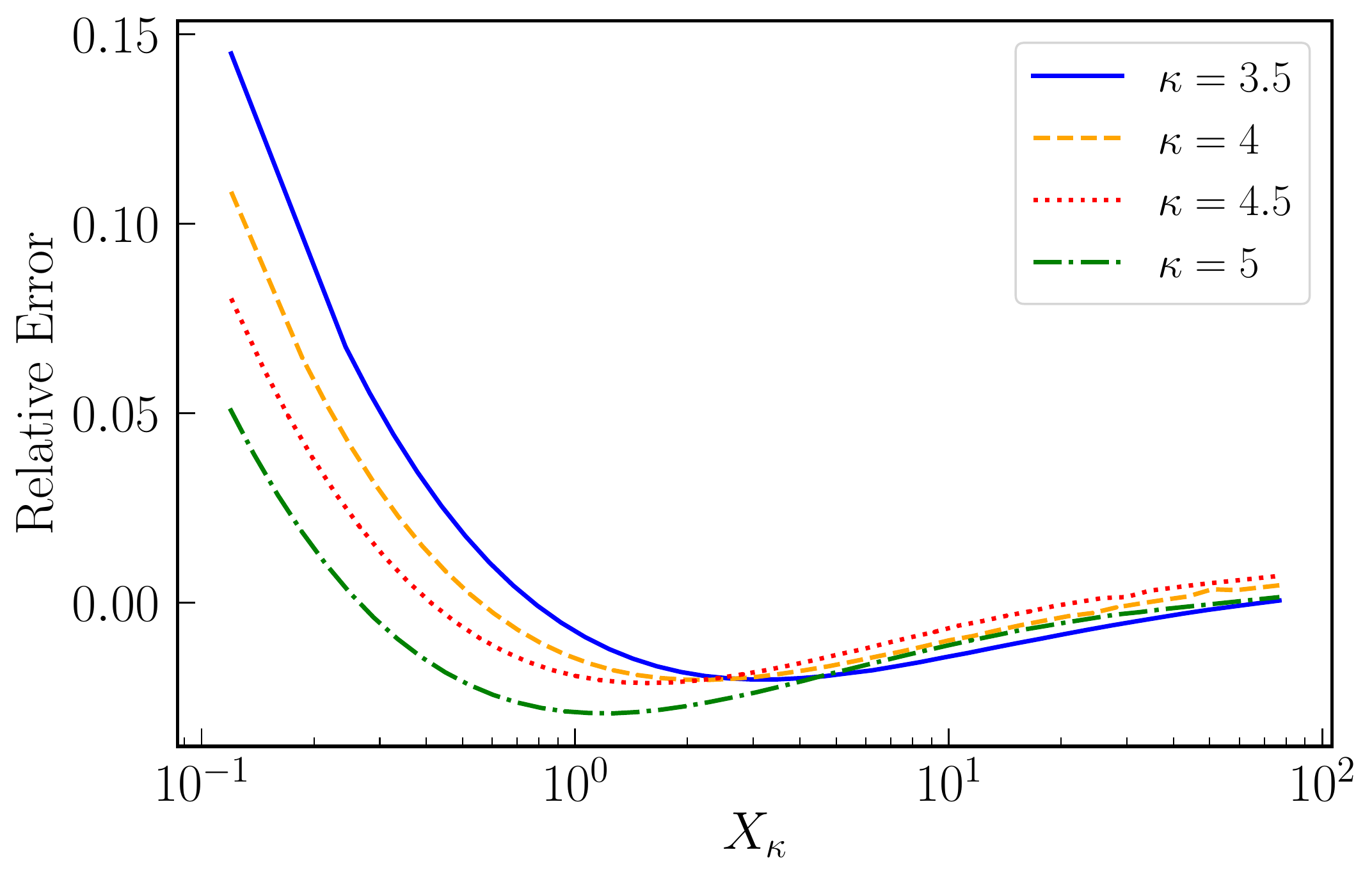}%
  \label{fig:2d}%
}
\caption{Comparison of our fits to \symphony{}'s numerically evaluated Faraday mixing coefficients. Panels (a) and (c) are log-log scale plots of $|\rho_Q|$ and $\rho_V$, respectively, versus $X_\kappa$ for various \kdf{} distributions.  The fitted values are shown by the solid line and  \symphony{}'s numerical evaluations are represented by the marks.  The sign change in $\rho_Q$ is represented in panel (a) by the change from "$+$" marks to "-" marks.  Relative error of the fits versus $X_\kappa$ on a log scale is shown in panels (b) and (d).  For these plots we set $w = 4$, $\theta = 60^{\circ}$, $B = 10$ Gauss, and varied $\nu$ to capture a range of $X_\kappa$.}
\label{fig:2}
\end{figure}
Also,
\begin{linenomath}
\begin{align}
\rho_V &=  \frac{2n_ee^2\nu_c\cos{\theta}}{mc\nu^2}\frac{K_0(w^{-1})}{K_2(w^{-1})}g(X_\kappa) \nonumber \\
    &\times \begin{cases}
     \frac{w^2+2w+1}{(25/8)w^2 + 4w+1}, & (\kappa = 3.5)\\
     \frac{w^2+54w+50}{(30/11)w^2 + 134w+50}, & (\kappa = 4)\\ 
     \frac{w^2+43w+38}{(7/3)w^2 + (185/2)w+38}, & (\kappa = 4.5)\\
     \frac{w+(13/14)}{2w+(13/14)}, & (\kappa = 5)
    \end{cases}
\end{align}
\end{linenomath}
where
\begin{equation}
\label{eqn:g_X}
g(X_\kappa) =  \begin{cases}
     1 - 0.17\ln{\left(1+0.447X_\kappa^{-1/2}\right)}, & (\kappa = 3.5)\\
     1 - 0.17\ln{\left(1+0.391X_\kappa^{-1/2}\right)}, & (\kappa = 4)\\ 
     1 - 0.17\ln{\left(1+0.348X_\kappa^{-1/2}\right)}, & (\kappa = 4.5)\\
     1 - 0.17\ln{\left(1+0.313X_\kappa^{-1/2}\right)}. & (\kappa = 5)
    \end{cases}
\end{equation}
These fits become inaccurate when $X_\kappa \lesssim 10^{-1}$ or when $\nu/\nu_c \lesssim 1$. \textcolor{black}{Fig. \ref{fig:2} compares these fitted rotativities to \symphony{}'s numerical evaluations.} 

\section{Conclusion}

We have corrected and extended earlier work on polarized radiative transfer coefficients.  In particular we have made sign corrections and corrected typographical errors in the emissivity and absorptivity fits (eqs. \ref{eqn:J_S_thermal}-\ref{eqn:x_A_S}) originally presented in P16 to be consistent with IEEE/IAU conventions and our own coordinate system.  In subsection \ref{Suscept_Tensor_Method} we present a new numerical integration method to calculate $\rho_Q$ from the components of the susceptibility tensor.  We find that combining the relevant components of the susceptibility tensor prior to integration dramatically reduces cancellation error in evaluation of $\rho_Q$.  Finally, we provide new fitting formulae for rotativities for various \kdf{} distributions in equations \ref{eqn:rho_Q_kappa}-\ref{eqn:g_X}.  \textcolor{black}{The updated fits for all coefficients are now available in \symphony{}\footnote{The current version is available at \url{https://github.com/AFD-Illinois/symphony}} and are implemented in the ray-tracing code \texttt{ipole}\footnote{The current version is available at \url{https://github.com/AFD-Illinois/ipole}}.} 

It is important to note that the corrections here do not affect the simulated images used in \citetalias{PaperVII} and \citetalias{PaperVIII}.  All images for these papers were run using the set of coefficients outlined in Appendix A of \citet{dexter_public_2016}.

{\color{black}
\appendix

\section{Definition of Emissivities and Absorptivities} \label{Appendix_A}

As shown in equations \ref{eqn:emissivityVector}-\ref{eqn:absorptivityVector}, calculations of emissivities and absorptivities from the radiative transfer equation require summing over harmonics, $n$, and performing a three-dimensional integral over momentum space.  The argument of the delta function in equations \ref{eqn:emissivityVector}-\ref{eqn:absorptivityVector} is
\begin{equation}
    y_n = \frac{n\nu_c}{\gamma} - \nu(1 - \beta\cos\xi\cos\theta).
\end{equation}
The integrand's dependence on the Stokes factor is given by
\begin{equation}
K_S = 
    \begin{cases}
     M^2J_n^2(z) + N^2J_n^{'2}(z), & \text{(Stokes I)}\\
     M^2J_n^2(z) - N^2J_n^{'2}(z), & \text{(Stokes Q)}\\ 
     0, & \text{(Stokes U)}\\
     -2MNJ_n(z)J_n^{'}(z), & \text{(Stokes V)}
    \end{cases}
    \label{eqn:K_S}
\end{equation}
where
\begin{equation}
    M = \frac{\cos\theta - \beta\cos\xi}{\sin\theta},
\end{equation}
\begin{equation}
    N = \beta\sin\xi,
\end{equation}
and
\begin{equation}
    z = \frac{\nu\gamma\beta\sin\theta\sin\xi}{\nu_c}.
\end{equation}
Here $J_n$ is a Bessel function of the first kind and $J'_n$ its derivative.  As shown in \ref{eqn:K_S}, $j_U$ and $\alpha_U$ are zero due to the symmetry of our coordinate system.

\section{Definition of Susceptibility Tensor} \label{Appendix_B}
Calculations of absorptivities and rotativities via the susceptibility tensor method require evaluation of specific components of the susceptibility tensor as shown in equations \ref{eqn:alphaSusceptibility}-\ref{eqn:rhoSusceptibility}.  In the following definition of the susceptibility tensor we assume that $\text{Im}(\omega) \ll \text{Re}(\omega)$.  The components of the susceptibility tensor are
\begin{equation}
    \chi_{ij} = \frac{2\pi i\omega^2_p}{\omega^2}\int^\infty_1 d\gamma (\gamma\beta)^3\frac{d\tilde{f}}{d\gamma}\mathbb{K}_{ij}(\gamma,\omega/\omega_c,\mathbf{k}),
\end{equation}
where $\omega \equiv 2\pi\nu$, $\omega_p^2 \equiv 4\pi nq^2/m$ is the species' plasma frequency, $q$ is the signed charge of the particle species (negative for electrons), $\omega_c \equiv qB/(mc)$, and 
\begin{equation}
    \label{eqn:Kernel_Suscept}
    \mathbb{K}_{ij}(\gamma,\omega/\omega_c,\mathbf{k}) = \int^\infty_0 d\tau e^{i\gamma\tau}\Phi_{ij}(\tau,\gamma.\omega/\omega_c,\mathbf{k}).
\end{equation}
Here,
\begin{equation}
    \Phi_{ij}(\tau,\gamma.\omega/\omega_c,\mathbf{k}) = \begin{pmatrix}
-\frac{1}{2}[\cos{(\frac{\omega_c}{\omega}\tau)}\it{I}_1(0) - \it{I}_1(2)] & -\frac{1}{2}\sin{(\frac{\omega_c}{\omega}\tau)}\it{I}_1(0) & -\cos{(\frac{\omega_c}{2\omega}\tau)}\it{I}_2(1)  \\
-\Phi_{12} & -\frac{1}{2}[\cos{(\frac{\omega_c}{\omega}\tau)}\it{I}_1(0) + \it{I}_1(2)] & \sin{(\frac{\omega_c}{2\omega}\tau)}\it{I}_2(1)   \\
\Phi_{13}  & -\Phi_{23}  & -\it{I}_3(0)  
\end{pmatrix},
\end{equation}
where
\begin{equation}
    \label{eqn:Analytic_I_1_0}
    \it{I}_1(0) = \frac{2((2\alpha^2+(\alpha^2-1)\delta^2+\delta^4)\sin{A}-(2\alpha^2-\delta^2)A\cos{A})}{A^5},
\end{equation}
\begin{equation}
    \it{I}_1(2) = -\frac{2\delta^2(3A\cos{A}+(A^2-3)\sin{A})}{A^5},
\end{equation}
\begin{equation}
    \it{I}_2(1) = \frac{2i\alpha\delta(3A\cos{A}+(A^2-3)\sin{A})}{A^5},
\end{equation}
\begin{equation}
    \label{eqn:Analytic_I_3_0}
    \it{I}_3(0) = \frac{6\alpha^2\cos{A}}{A^4} - \frac{2\cos{A}}{A^2} + \frac{6\delta^2\sin{A}}{A^5} - \frac{4\sin{A}}{A^3} +\frac{2\alpha^2\sin{A}}{A^3},
\end{equation}
\begin{equation}
    \alpha = \gamma\beta\tau\cos{(\theta)},
\end{equation}
\begin{equation}
    \delta = \frac{2\gamma\beta\omega\sin{(\theta)}}{\omega_c}\sin{(\frac{\omega_c\tau}{2\omega})},
\end{equation}
and
\begin{equation}
    A = \sqrt{\alpha^2+\delta^2}.
\end{equation}
The integral in Equation \ref{eqn:Kernel_Suscept} is convergent for real $\omega$ except when $\cos{(\theta)} = 0$ (Section 2 and Section 3 of P18 consider the general case of complex $\omega$).  
}

\acknowledgments
This work was supported by National Science Foundation grant OISE 17-43747 and by a Donald C. and F. Shirley Jones Fellowship to G.N.W. AP is supported by the National Science Foundation (NSF) Graduate Research Fellowship Program under Grant No. DGE1656466.

\bibliography{Bib, transfer_coefficients, EHTC}
\bibliographystyle{aasjournal}

\end{document}